\documentclass[12pt]{article}
\usepackage{amsmath,amssymb
}
\begin{document}

\begin{center}
{\large {\bf Discussion: Remarks on ``Random Sequences''}}
\end{center}

\begin{center}
 Branden Fitelson and Daniel Osherson \\[6pt]
 
 
\end{center}

\thispagestyle{empty}

\section{Setup}

We consider evidence relevant to  whether a (possibly idealized) physical process is producing its output randomly.   For definiteness, we'll consider a coin-flipper $\mathcal{C}$ which reports ``$\mathtt{H}$'' for a heads  and ``$\mathtt{T}$'' for a tails.  By $\mathcal{C}$ producing its output ``randomly,'' we mean $\mathtt{H}$ and $\mathtt{T}$ have equal probability and trials are independent.  If $\mathcal{C}$ produces its output randomly (in the above sense), then we'll say that $\mathcal{C}$ is a \textsl{random device}.

There are many potential reasons to believe that $\mathcal{C}$ is a random device (or that it's not).  We might know something about its manufacture, or be told that $\mathcal{C}$ is random (or not) on good authority, \emph{etc.}  Our question is whether there is any information in $\mathcal{C}$'s \emph{output} that bears on whether  $\mathcal{C}$ is random.  We'll consider two statistics (concerning output sequences generated by $\mathcal{C}$) that are often taken to provide information about $\mathcal{C}$'s randomness.  The first statistic is the number of runs in an output sequence \cite{ww}.  The second statistic is the number of heads \emph{versus} tails.  

In order to evaluate the utility of these statistical tests for randomness, we'll focus on the following two potential output sequences:
\begin{itemize}
\item[($A$)] $\mathtt{HTTHTHHHT}$
\item[($B$)] $\mathtt{HHHHHTTTT}$
\end{itemize}

  A \textsl{run} in a sequence is a maximal non-empty segment consisting of adjacent equal elements.  For example ($A$) has six runs whereas ($B$) has just two.  If $\mathtt{H}$s and $\mathtt{T}$s alternate randomly then the number of runs after $N$ trials is a random variable whose cumulative distribution is given by counting the number of sequences of length $N$ with $r$ or fewer runs (or conversely, $r$ or greater runs).   Doing the relevant calculations for ($A$), we deduce: If $\mathcal{C}$ is a random device then the probability is  $0.363$ of producing these many (\emph{viz.}, six) runs or more in a sequence of length nine.  Because of this,  advocates of the runs test say that producing ($A$) does \emph{not} strongly disconfirm that $\mathcal{C}$ is a random device.  For ($B$) the same calculations imply: If $\mathcal{C}$ is a random device then the probability is $0.035$ of producing this many runs (\emph{viz.}, two) or fewer in a sequence of length nine.  In this case, advocates of of the runs test say that $\mathcal{C}$'s generating ($B$) \emph{does} strongly disconfirm  $\mathcal{C}$'s randomness.\footnote{\label{royall}Standard objections to evidential interpretations of classical statistical tests have been recently surveyed in \cite{royall} and \cite{greco}.  Our objection will be somewhat different from earlier concerns.}
    
  The binomial test gives the probability of throwing at least $x$ heads in $n$ tosses of the coin (or the probability of throwing at most $x$ heads if they are fewer than $\frac{n}{2}$).  In both ($A$) and ($B$), we see 5 heads in 9 tosses. We compute that if $\mathcal{C}$ is a random device, then producing a sequence with five or more heads has probability $0.5$. Because of this,  advocates of the binomial test say that the fact that $\mathcal{C}$ generates either sequence does \emph{not} strongly disconfirm the claim that $\mathcal{C}$ is a random device.
  
  We've exploited two statistical tests to evaluate evidence regarding whether $\mathcal{C}$ is a random device. If ($B$) is the output, the first test (``runs'') classifies this as strongly disconfirmatory of $\mathcal{C}$'s randomness.  If the output is ($A$), the first test does not deem this to be strongly disconfirmatory. The second test (``binomial'') views neither case ($A$) nor ($B$) as constituting strong evidence against $\mathcal{C}$'s randomness. While these tests may disagree with each other, they each seem  to be perfectly self-consistent.  But, there is a problem \ldots
  
\section{The Problem}

At a given position of the sequence produced by $\mathcal{C}$, there are more potential events than just ``heads'' and ``tails.''  For example, let $X = \{1,4,9\}$, and define:
\begin{itemize}
\item  Position $i$ of $\mathcal{C}$'s output holds a \textsl{hail} ($\mathtt{h}$) iff either $i \in X$ and position $i$ holds a head ($\mathtt{H}$), or $i \notin X$ and position $i$ holds a tail ($\mathtt{T}$).

\item Position $i$ of $\mathcal{C}$'s output holds a \textsl{tead} ($\mathtt{t}$) iff either $i \in X$ and position $i$ holds a tail ($\mathtt{T}$), or $i \notin X$ and position $i$ holds a head ($\mathtt{H}$).
\end{itemize}

\noindent Given these definitions of teads and hails, we see that $\mathcal{C}$ generates ($A$) iff $\mathcal{C}$ generates ($a$), and $\mathcal{C}$ generates ($B$) iff $\mathcal{C}$ generates ($b$).\begin{center}
\renewcommand{\arraystretch}{1.5}
$\begin{array}{|ll|ll|}\hline
(A) & \mathtt{HTTHTHHHT} & (a) & \mathtt{hhhhhtttt}\\
(B) & \mathtt{HHHHHTTTT} & (b) & \mathtt{htththhht}\\\hline
\end{array}$
\end{center}

 Let us respond at once to the
concern that teads and hails are ``unnatural,'' ``position dependent,'' or otherwise ``gerrymandered.''  Such characterizations seem no more applicable to teads/hails than to
heads/tails. For, we  have the following symmetry:\begin{itemize}
\item Position $i$ of $\mathcal{C}$'s output holds a tail ($\mathtt{T}$)  iff either $i\in X$ and
position $i$ holds a tead ($\mathtt{t}$), or $i\not\in X$ and position $i$ holds a hail ($\mathtt{h}$).

\item Position $i$ of $\mathcal{C}$'s output holds a head ($\mathtt{H}$)  iff either $i\in X$ and
position $i$ holds a hail ($\mathtt{h}$), or $i\not\in X$ and position $i$ holds a tead ($\mathtt{t}$).
\end{itemize}

\noindent  Someone who thinks in terms of heads/tails may well find teads/hails to be ``derivative.''
But someone who thinks in terms of teads/hails will make the parallel claim about heads/tails.
It's not obvious how to break the symmetry.  Moreover, $\mathcal{C}$ produces an unbiased, independent sequence of heads/tails iff $\mathcal{C}$ produces an unbiased, independent sequence of teads/hails. (This is easy to verify.)  Therefore, the runs test applied to teads/hails is as relevant to the randomness of
$\mathcal{C}$ as the runs test applied to heads/tails.  Unfortunately, applying the runs test to teads/hails  leads to a \emph{reversal} of our initial assessment (in terms of heads/tails).

To see this, just count the number of runs of teads/hails in ($a$) and ($b$), above.  We see that ($a$) has two runs and ($b$) has six.  Doing the relevant calculations for ($a$), we deduce: If $\mathcal{C}$ is a random device then the probability is $0.035$ of producing a sequence (of length nine) with so few $\mathtt{t}$/$\mathtt{h}$ runs.  The advocate of the runs test should say that this constitutes strong evidence against the claim that $\mathcal{C}$ is a random device.  And, for ($b$), we deduce: If $\mathcal{C}$ is a random device then the probability is  $0.363$ of producing a sequence (of length nine) with so many $\mathtt{t}$/$\mathtt{h}$ runs.  The advocate of the runs test should say that that this does \emph{not} constitute strong evidence against the claim that $\mathcal{C}$ is a random device.  Thus, the use of teads/hails instead of heads/tails reverses the evidential verdict implied by the runs test!

Underlying this phenomenon is alteration of the ``rejection set'' in the passage
from heads/tails to teads/hails. The rejection set is composed of the sequences
whose number of runs is ``too extreme'' to be easily compatible with $\mathcal{C}$'s randomness.
A given, potential output from $\mathcal{C}$ might be considered extreme when the rejection
set is reckoned in terms of runs of heads/tails but not teads/hails, and conversely.
So the runs test is ambiguous unless some reason can be given to favor one way
of counting runs over all the competing ways (and finding such a reason seems problematic).

The same sort of reversal can be achieved for the binomial test as well.  To wit, consider the following pair of potential outcome sequences:
\begin{itemize}
\item[($A$)] $\mathtt{HTTHTHHHT}$
\item[($D$)] $\mathtt{TTTTTTTTT}$
\end{itemize}
Then, let $Y = \{2,3,5,9\}$, and define:
\begin{itemize}
\item Position $i$ of $\mathcal{C}$'s output holds a \textsl{schmail} ($\mathtt{t}$) iff either $i \in Y$ and
position $i$ holds a tail ($\mathtt{T}$), or $i \notin Y$ and position $i$ holds a head ($\mathtt{H}$).

\item Position $i$ of $\mathcal{C}$'s output holds a \textsl{schmead} ($\mathtt{h}$) iff either $i \in Y$ and
position $i$ holds a head ($\mathtt{H}$), or $i \notin Y$ and position $i$ holds a tail ($\mathtt{T}$).
\end{itemize}

\noindent Similarly to before, $\mathcal{C}$ is a random device for generating heads/tails iff $\mathcal{C}$ is a random device for generating schmails/schmeads. But, $\mathcal{C}$ produces ($A$) or ($D$) iff  $\mathcal{C}$   produces ($c$) or ($d$), respectively.  So we can apply the binomial test to both (pairs of) sequences of events:\begin{center}
\renewcommand{\arraystretch}{1.5}
$\begin{array}{|ll|ll|}\hline
(A) & \mathtt{HTTHTHHHT} & (c) & \mathtt{ttttttttt}\\
(D) & \mathtt{TTTTTTTTT} & (d) & \mathtt{htththhht}\\\hline
\end{array}$
\end{center}
\noindent Applying the binomial test to the schmeads and schmails in ($c$) yields: If $\mathcal{C}$ is a random device then the probability is $0.004$ of producing an event with so few $\mathtt{h}$s.  The advocate of the binomial test should therefore view $\mathcal{C}$'s generating ($c$) as strong evidence against the claim that $\mathcal{C}$ is a random device.   We saw earlier that the binomial test does \emph{not} imply that ($A$) is an improbable sequence if  generated randomly. As such, advocates of the binomial test should not view $\mathcal{C}$'s generation of ($A$) as strong evidence against $\mathcal{C}$'s randomness.  The same reversal affects ($D$) and ($d$).  Once again, the test's implications about evidential relevance depend on which concepts we employ.\footnote{Such reversals will plague \emph{any} statistical test for randomness that we have encountered (see \cite[Ch.~2]{lv} for a recent survey).}

\section{What the Problem is \emph{Not}}
 
The teads/hails terminology resonates with Goodman's \cite[Ch.~3]{goodman} use of grue/bleen
to question the basis of projections to the future. But this is not the point
of the present discussion. Indeed, whether one reckons an output sequence as
\texttt{HTTHTHHHT} versus \texttt{hhhhhtttt} has no bearing on predictions about the next coin
toss. After the 9th output, heads are invariably teads and tails hails. So if
you expect a head [tail] there is no harm in announcing a tead [hail]. The
situation is thus different from Goodman's since projecting the greenness of
emeralds ultimately conflicts with projecting grueness (after time $t$ the
two kinds of emeralds look different). The same remarks apply to schmeads and
schmails.
 
In contrast, the choice between heads/tails versus teads/hails appears to
alter the verdict of standard statistical tests about the here and now,
namely, whether $\mathcal{C}$ is producing its output randomly. Driving the ambiguity
is the fact that \emph{$\mathcal{C}$ issues heads and tails in a uniform, independent way
just in case the same is true for teads and hails}, hence, the tests apply
equally in the two cases. Preserving the ``null hypothesis'' of uniformity
and independence across shifts in vocabulary is not a feature of the
grue/bleen puzzle.\footnote{In this sense, the present phenomenon is perhaps more similar to Miller's \cite[Ch.~11]{miller} language-dependencies than Goodman's.}
 
Of course, at a more abstract level, both teads/hails and grue/bleen point to
the language dependence of inductive inference. If we denoted both heads and
tails by \textit{theds} without specialized vocabulary for each then we might
be struck by the fact that $\mathcal{C}$ produces nothing but theds. But our point is
more specific. Standard statistical tests for the randomnesss of $\mathcal{C}$ yield
conflicting results even though $\mathcal{C}$ is random with respect to one vocabulary
if and only if it is random with respect to the other. Unless a principled
choice can be made among candidate vocabularies, the tests are bound to offer
equivocal verdicts.
 
Embracing the language dependence of the tests, moreover, does not seem to
be a viable response to the ambiguity. It makes no sense to declare different
levels of confidence for $\mathcal{C}$'s being random ``in the sense of heads/tails''
compared to $\mathcal{C}$'s being random ``in the sense of teads/hails.'' For (to
repeat), \emph{$\mathcal{C}$ is random in one sense if and only if it is random in the other}.
A better response, it seems to us, is to abandon the tests altogether, along
with any other attempt to harness $\mathcal{C}$'s output to compute its likelihood
assuming randomness within a null hypothesis framework.

\section{Lessons Learned}

What is the value of a statistical test whose outcome
is so sensitive to the concepts used to describe the data?  It would appear that this kind of null hypothesis testing in the service of evaluating the randomness of $\mathcal{C}$ is of little epistemic value.  Indeed, it is often noted that all sequences of a given length have the same probability
of being generated by an unbiased independent source. So there's no such thing as an ``atypical'' sequence that is ``unlikely'' to be generated if $\mathcal{C}$ is random.  \emph{All sequences are atypical, surprising, coincidental, etc.}\footnote{Ironically, some advocates of statistical tests for randomness \cite[\S 1.8.1]{lv} seem to think that this reveals a \emph{shortcoming} of purely probabilistic  assessments of output sequences.  On the contrary, we think the present considerations show that there is something wrongheaded about any approach to randomness that appeals solely to properties of output sequences.} 

Yet, intuitively, it seems reasonable (in some sense) to be sceptical about the randomness of a source that relentlessly produces heads.  What explanation can we offer for such doubt?  Prior to seeing any output, there are many alternatives to the hypothesis that $\mathcal{C}$ is random.  One alternative is that \emph{a human mind controls the output}.  The human-control hypothesis enjoys a relatively elevated prior probability because there are so many human minds in the neighborhood.  (If we lived far away, we might be surrounded by teads/hails speakers, leading to different priors about the character of $\mathcal{C}$.)   The likelihood of a long initial stretch of heads --- \emph{given human control} --- is relatively high (simply because that's the kind of thing a human would do), so the posterior probability of human-control comes to swamp the priors.

On our view, belief that $\mathcal{C}$  is random should not be based solely on $\mathcal{C}$'s output.  Ideally, it is inspection of $\mathcal{C}$'s mechanism that grounds convictions about randomness (perhaps because of symmetries discovered, or for deeper reasons involving quantum theory, \emph{etc.}).  On this view, a sequence of events is ``random'' iff it has been generated by a random device.

\end{document}